\documentclass[twocolumn, longbibliography, floatfix]{revtex4-2}
\usepackage{graphicx}
\usepackage{amsmath}
\usepackage{xcolor}
\usepackage{amssymb}
\usepackage{siunitx}
\usepackage{float}
\usepackage{multirow}
\usepackage{comment}
\usepackage{array}
\usepackage{booktabs}

\usepackage{placeins}

\usepackage{pgfplots}

\usepackage{enumitem} 

\usepackage{hyperref}
\usepackage{orcidlink}

\newcommand{\new}{\color{red}}
\renewcommand{\new}{} 

\begin{document}
	
	\title{Data-driven synthesis of high-fidelity triaxial magnetic waveforms for quantum control}
	
	\author{Giuseppe Bevilacqua
	\orcidlink{0000-0002-7298-6185}}
	\affiliation{DSFTA, Università di Siena, Via Roma 56, 53100 Siena, Italy}
	
	\author{Valerio Biancalana
	\orcidlink{0000-0002-1642-5391}}
	\email{email: valerio.biancalana@unisi.it}
	\affiliation{DSFTA, Università di Siena, Via Roma 56, 53100 Siena, Italy}
	
	\author{Roberto Cecchi
	\orcidlink{0000-0001-9091-2780}}
	\affiliation{DSFTA, Università di Siena, Via Roma 56, 53100 Siena, Italy}
	
	\date{\today}

	\begin{abstract}
		We present a system for generating arbitrary, triaxial magnetic waveforms with a spectral content spanning from DC to tens of kHz, a critical capability for quantum control and spin manipulation. To compensate for amplifier-coil dynamics, we implement a data-driven approach to identify a numerical compensation model. The method parametrizes the system response using a Finite Impulse Response (FIR) filter calibrated on the specific waveform to be generated. The application of a driving signal designed via frequency-domain inversion of the identified model enables the synthesis of complex field sequences with sharp transitions between static and single- or multi-frequency temporal segments. The work is validated with experimental results demonstrating waveform fidelity and transient performance, thereby showcasing the precision and feasibility of the method.
	\end{abstract}
	
	\maketitle

\section{Introduction}
\label{sec:introduction}
Precise control of time-dependent, triaxial magnetic fields is a recurring requirement in modern atomic physics experiments. The ability to apply magnetic fields with arbitrary temporal profiles along three orthogonal axes enables coherent quantum-state control and the implementation of complex dynamical protocols. Such capabilities are essential, for instance, in the coherent manipulation of spin and pseudospin, in the engineering of effective Hamiltonians, and in the realization of time-dependent Zeeman or synthetic spin--orbit couplings
\cite{kestner_prl_13,senko_science_14,glaser_epjd_15,martin_prx_17,bevilacqua_pra_22,fregosi_sr_23,petiziol_epl_24,fregosi_sr_26}.

Arbitrary waveform generation is equally relevant for the active compensation of environmental magnetic disturbances, where adaptive {\new feedback and} feed-forward strategies are particularly effective
\cite{bevilacqua_prappl_19,odwyer_rsi_20,pyragius_rsi_21,ponikvar_electronics_23}%
{\new , and feedback combined with feed-forward has brought the stability of a coil current and the rejection of mains-synchronous ambient fields to the sub-ppm level \cite{merkel_rsi_19}
}.

In all these applications, the field must cover a spectral range from DC to several tens of kilohertz while maintaining amplitude accuracy and phase coherence across the three axes --- a combination of requirements that places demanding constraints on the current-driving electronics \cite{robinson_neuroimage_22}. The central challenge is that the amplifier and coil act as a filter, whose characteristics are only approximately known. {\new In particular, the inductance of the coils and the finite bandwidth and compliance of the amplifiers distort the programmed current, especially at sharp transitions between static and oscillating segments.}

{\new Compensating the dynamics of a control chain by pre-distorting the drive signal is well established outside atomic physics. In magnetic resonance imaging, the response of gradient coils perturbed by eddy currents is corrected by inverse filtering, with the filter derived from an analytical model \cite{ahn_ieee_91}. In solid-state quantum processors, the response of the control line is measured, typically using the qubit itself as a probe, and used to design digital pre-distortion filters (direct inversion of the transfer function, or real-time IIR/FIR filters) for microwave, flux and baseband pulses on nanosecond timescales \cite{gustavvson_prl_13,rol_apl_20,duan_prl_26}. Alternatively, instrumental distortions can be included in the optimal-control design, since inverses of instrument responses may be ill-defined or unstable \cite{rasulov_jpc_25}.
}

{\new In atomic physics, instead, active field control has mostly addressed noise and drift rejection. Where the response of the current-control loop had to be compensated, this was done with an analogue stage approximating the inverse of the measured transfer function, to be adapted to each power supply and coil \cite{merkel_rsi_19}. To the best of our knowledge, a data-driven, software-based and rapidly reconfigurable pre-compensation for the synthesis of arbitrary triaxial waveforms with high-inductance coils has not been reported.}

We present an apparatus for generating arbitrary magnetic field waveforms over the aforementioned frequency range. The method learns the filter's behavior from data and then precompensates the input signal so the output magnetic field matches the target. The system response is modeled by a Finite Impulse Response (FIR) filter whose coefficients are evaluated by fitting measured input--output data \cite{unser_sp_94}. Once the filter is identified, its inversion yields the driving voltage waveform required to produce any user-defined field profile. The proposed methodology embodies a data-driven procedure that avoids reliance on accurate circuit models and specifications and captures unmodeled dynamics through experimental identification, providing a flexible and robust route to waveform synthesis, with enhanced fidelity in selected critical time intervals.

The resulting voltage signals, synthesized by a digital-to-analog converter (DAC) are applied to a set of three power amplifiers driving independent perpendicular coils, thus producing the corresponding magnetic field components. This approach enables direct and flexible software control over the temporal evolution of each component, ensuring an accurate correspondence between the designed and realized field waveforms and synchronicity among the generated field components.

{\new Besides the specific regime considered here (DC to tens of kHz, high-inductance coils driven by power amplifiers through a reconfigurable compensation network, three synchronous axes), the novelty of this work, with respect to the approaches above, lies in a combination of elements. (i) The identification relies on the monitored coil current and on a calibration signal derived from the target waveform itself, without a quantum probe or dedicated step-response measurements, so that the excitation covers the spectrum actually required. (ii) The FIR coefficients are obtained by an optimisation that concentrates the accuracy on user-defined critical intervals. (iii) The method is validated experimentally against both a nominal-circuit model and an FFT-based characterisation. (iv) Recalibration after a hardware or waveform change takes only a few tens of seconds.}

The remainder of the paper is organized as follows. Section~\ref{sec:setup} describes the hardware and the identification--inversion procedure. Section~\ref{sec:results} presents experimental results for a waveform representative of a spin-manipulation protocol. Section~\ref{sec:discussion} discusses the performance, limitations, and practical advantages of the methodology, and Sec.~\ref{sec:conclusions} draws the conclusions.

	\section{Setup}
	\label{sec:setup}

	\subsection{Impedance Compensation and Frequency Response Considerations}
	\label{subsec:impedance}
	
The inductive impedance of a coil grows linearly with frequency, progressively limiting the current that an amplifier with a fixed compliance voltage can deliver. Without countermeasures, this restricts the usable bandwidth, reducing the field amplitude achievable at higher frequencies.
	
	\begin{figure}[ht!]
		\includegraphics[width=\columnwidth]{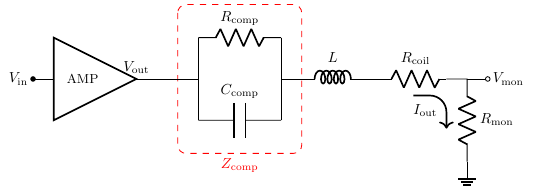}
		\caption{Schematics of the load and the compensation network, featuring a parallel $R_\mathrm{comp}$–$C_\mathrm{comp}$ compensation network in series with the coil $L_\mathrm{coil}$ and the monitoring resistor $R_\mathrm{mon}$.
			The parallel branch compensates for the coil’s inductance: the capacitive reactance enhances the maximum attainable current near the resonant frequency, while the resistive term ensures proper DC operation.
			The low-side monitor resistor enables the measurement of the actual current $I_\mathrm{out}$ through the voltage $V_\mathrm{mon}$ and hence an estimate of the magnetic field, provided that accurate calibration factors~\cite{bevilacqua_apdap_25} are available. The physical implementation of this network, with selectable components, is shown in the amplifier schematic of Fig.~\ref{fig:schemacircuito}.}
		\label{fig:amplifier_compensation}
	\end{figure}

A standard remedy is to insert a compensation network that partially cancels the coil reactance, as shown in Fig.~\ref{fig:amplifier_compensation}. Depending on the choice of components, such a network can flatten the overall transfer function or boost the available current within selected frequency ranges.
	
The compensation network modifies the effective load seen by the amplifier, introducing a non-trivial frequency dependence in both amplitude and phase that must be accounted for when designing the driving signal.

The system operates in open-loop, feed-forward mode. A model-based approach would design the pre-compensated driving signal analytically from knowledge of the amplifier and load~{\new\cite{ahn_ieee_91}}, but requires accurate component values and experimental validation to handle parasitics and tolerances. The implemented data-driven procedure instead infers the system transfer function directly from measurements, making component uncertainties irrelevant and enabling rapid reconfiguration when the hardware changes. The procedure relies entirely on standard numerical filtering and least-squares routines available in common laboratory software (LabVIEW in our case), making it straightforward to implement and easy to adapt to different experimental setups.

	\subsection{Hardware}
	\label{subsec:hardware}
	
The custom amplifier described below implements the compensation principle of Sec.~\ref{subsec:impedance}. Its impedance network is reconfigurable to accommodate coils with different parameters or to meet the specifications of the field waveforms to be generated. The presented setup utilizes several coils in various arrangements with diverse specifications 
	(resistances of \SIrange{2}{50}{\ohm}, inductances of \SIrange{0.5}{10}{mH}, and 
	coil constants of \SIrange{100}{2000}{nT/mA}) supplied by variously configured power amplifiers. 
	
	\begin{figure*}
		\centering
		\includegraphics[width=0.9\linewidth]{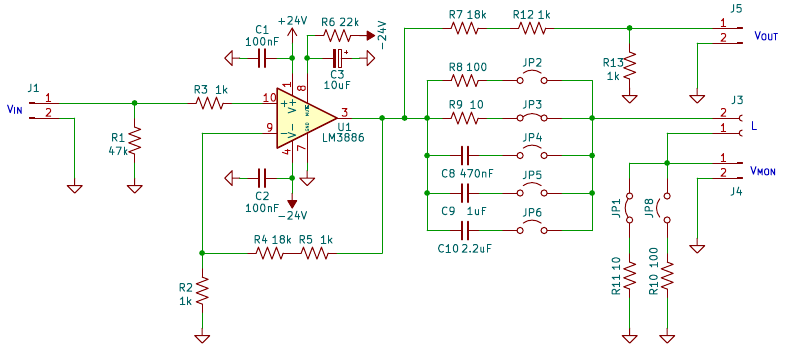}
		\caption{The amplifier is built around an LM3886 integrated circuit, its design employs a linear non-inverting topology with full DC-coupling to maintain signal integrity from input to output. It features a reconfigurable conditioning network, adjustable via jumpers, allowing the system to adapt its response and frequency shaping to the specific requirements of the inductive load or of the designed waveforms.}
		\label{fig:schemacircuito}
	\end{figure*}
	
	The amplifier schematic (Fig.~\ref{fig:schemacircuito}) features an LM3886 power operational amplifier, configured in a non-inverting topology and powered by a $\pm$\SI{24}{V} dual supply. The input signal is fed to the non-inverting terminal through a conditioning network consisting of resistors that define the input impedance and ensure DC coupling. The feedback path is implemented via a resistive network composed of $R_4$, $R_5$, and $R_2$, which sets the overall voltage gain to 20.
	
	The amplifier drives the coil connected to the output through a network of resistors and capacitors, selectable via jumpers. The capacitors $C_8$, $C_9$, and $C_{10}$ can be engaged to modify the equivalent impedance seen by the output, thereby mitigating the effects of the load's inductive reactance. These elements introduce an equalization behavior that shapes the frequency response of the system within a specific frequency range. The overall system response can be rapidly and coarsely reconfigured to meet current experimental requirements.
	
	Resistors $R_8$ and $R_9$, placed in parallel with the capacitors, provide a path for DC current flow through the coil. Furthermore, $R_{10}$ and $R_{11}$ serve as a shunt resistance for current monitoring. The voltage drop across these resistors is proportional to the delivered current, allowing for an indirect measurement of its instantaneous value.
	
	Supply bypass capacitors are included to reduce high-frequency impedance and improve the overall stability of the amplifier. Finally, diodes and LEDs complete the power stage, providing protection and visual indication of the supply rails.
	
	A NI-DAQ card generating three independently programmable waveforms sampled at \SI{125}{kHz} provides the input signals for three amplifiers, each configured for the respective coil specifications. {\new The relevant specifications of the DAQ system are summarized in Tab.\ref{tab:DAQspec}}

	\begin{table}[H]
		\new
		\centering

		\begin{tabular}{lll}
			\hline \hline
			& Description & Value \\
			\hline
			\multirow{5}{*}{DAC} & Number of channels & 4 \\
			& Voltage range used & $\pm 1$\,V \\
			& Resolution & 16\,bits \\
			& Max update rate & 719\,kS/s \\
			& Timing accuracy & $50\,\mathrm{ppm} \times \Delta t_\mathrm{sample}$\\
			\hline
			\multirow{5}{*}{ADC} & Number of channels & 4 \\
			& Voltage range used & $\pm 1$\,V \\
			& Resolution & 16\,bits \\
			& Max sample rate (4 ch) & 125\,kS/s \\
			& Timing accuracy & $50\,\mathrm{ppm} \times \Delta t_\mathrm{sample}$\\
			\hline \hline
		\end{tabular}
				
		\caption{Relevant specifications of the NI USB-6343 DAQ system used in this work, for the configuration adopted in the experiment.}
		\label{tab:DAQspec}
	\end{table}
	
	While the hardware allows for coarse impedance matching via jumpers, the remaining fine-grained dynamics and non-idealities are addressed through the numerical identification procedure described in the following section.

	\subsection{System identification and waveform assignment}
	\label{subsec:identification}
	
	\begin{figure}[htbp]
		\centering
		\includegraphics[width=\columnwidth]{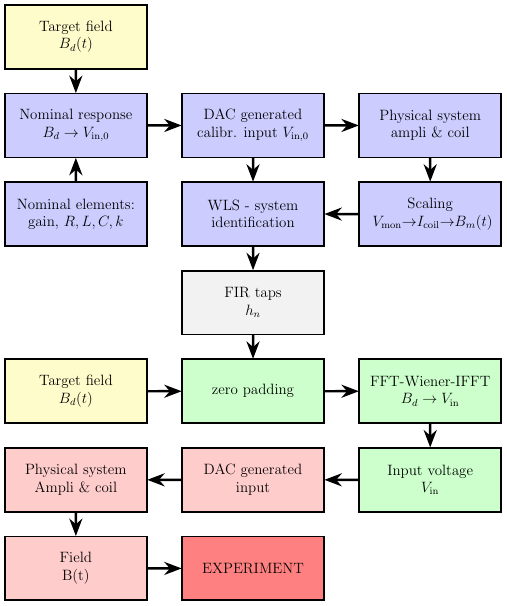}
		\caption{Workflow for system identification and waveform pre-compensation. 
			The procedure is divided into two main stages: 
			Phase 1 (in blue: identification). A calibration voltage $V_\mathrm{in,0}$, evaluated from an ideal behavior, excites the system. The resulting coil current $I_{coil}$ is measured via a shunt resistor ($R_\mathrm{mon}$), and the magnetic field $B(t)$ is inferred through the coil constant $k$. A FIR model is then identified using a WLS algorithm, which allows for temporal weighting to exclude initial transients and enhance the accuracy in selected critical segments.
			Phase 2: (in green: assignment). The identified FIR taps ($\hat{h}$) are used to perform a frequency-domain inversion. The target field $B_d(t)$ is processed through a Wiener-regularized deconvolution to derive the final pre-compensated input voltage $V_\mathrm{in}(t)$, ensuring numerical stability even in the presence of non-minimum phase zeros or spectral nulls. This signal excites the system and provides the high-fidelity field needed in the experiment.}
		\label{fig:flowchart_dpd}
	\end{figure}

The identification--inversion procedure consists of two steps (see Fig.~\ref{fig:flowchart_dpd}): (i) identifying a model of the system's transfer function from measured data, and (ii) inverting that model to compute the driving voltage $V_\mathrm{in}(t)$ that shall produce the desired field $B_d(t)$. We employ a hybrid approach in which identification is performed in the time domain and the inversion in the frequency domain.
	
The amplifier--coil system is treated as linear and time-invariant (LTI)~\cite{ljung_book_99}, an assumption well justified by the operating conditions: the amplifier works within its linear range, there are no ferromagnetic materials near the coils, and operation frequencies are low enough to make parasitic capacitances and radiation losses negligible. Under these conditions the instantaneous field is proportional to the instantaneous coil current, which is inferred from the voltage across the monitor resistor. The system is first driven with a calibration voltage $V_\mathrm{in,0}(t)$, obtained from the target waveform $B_d(t)$ by a simple FFT-based inversion of the nominal passive network (the amplifier gain is assumed frequency-independent at this stage).
	
A non-parametric FFT-based identification --dividing the output spectrum by the input spectrum--is the simplest option~{\new\cite{gustavvson_prl_13}} and gives reasonable results, as shown in Sec.~\ref{sec:results}.
	
Better results are obtained instead with a parametric, time-domain identification based on a Finite Impulse Response (FIR) filter~\cite{haykin_book_96,ljung_book_99}. The FIR taps $h_n$ are determined by minimising $\|F_\mathrm{FIR}(V_\mathrm{in,0})-B_m\|_W^2$ over the measured calibration data~{\new\cite{rol_apl_20, duan_prl_26}}. As this is a convex problem~\cite{ljung_book_99} it can be solved reliably and efficiently by weighted least squares (WLS) optimization. The use of a calibration signal $V_\mathrm{in,0}(t)$ that (even if roughly) approximates the required one ensures that the recorded data contain the features relevant for a successful and accurate identification.
	
The FIR/WLS approach offers two advantages over the FFT alternative that are directly relevant to atomic-physics applications: it is causal and parametric (the filter has a well-defined impulse response that can be stably inverted), and it allows user-defined temporal weights to be assigned to different segments of the signal.
	
The weighting is a relevant feature for experimental use: the operator specifies which time intervals are critical (e.g., the transition from the optical-pumping phase to the measurement phase), and the WLS fit concentrates the model accuracy there while tolerating larger residuals elsewhere. In the results reported below, a weight 30 times larger was assigned to the \SI{10}{ms} interval around the $t=0$ transition; the results were found not to be sensitive to moderate variations of this value.

An IIR model would require fewer parameters to capture long-lived transients, but its identification leads to a non-convex optimisation problem and  
the resulting filter may be unstable;{\new  cascaded first-order IIR sections have nevertheless been used for slow exponential tails \cite{rol_apl_20}.}

The FIR structure avoids both issues at the cost of a larger number $N$ of taps. In our implementation $N$ is chosen by the operator by inspecting the tap values and verifying that they decay to zero well before the $N^{th}$ tap (see Fig.~\ref{fig:fircoefficients}a). This is easily assessed visually and takes only a few calibration runs. For the results reported here, $N=120$ was sufficient. An automatic selection would be feasible as well, but was not in the needs of our experiment.

Once the FIR model is identified, the driving voltage is obtained by inverting it. Direct time-domain inversion is unreliable when the system has non-minimum-phase zeros~{\new\cite{rasulov_jpc_25, duan_prl_26}}, and we verified that an approximate FIR inverse works only for very simple loads.

A method based on frequency-domain inversion is more robust~\cite{oppenheim_book_99}. 
{\new Zero-padding ensures linear (non-circular) convolution when the inverse
filter is applied, while the Wiener regularisation prevents the inverse filter
from diverging where the system response is small and limits its effective
duration, thus avoiding wrap-around~\cite{kirkeby_ieee_98}.}	
In our implementation the regularisation parameter is chosen empirically to balance noise suppression against bandwidth.

	\section{Results}
	\label{sec:results}
The identification algorithm was first validated through simulations (not shown for brevity), enabling the system to be tested across a wide and controlled range of operating conditions to provide significant reference data. The system was then implemented as a three-axis field generator and tested experimentally. To demonstrate the method, we defined a test target, $B_d(t)$, which is representative of a typical spin-manipulation protocol.
	
	\subsection{Required current waveforms}
	\label{subsec:esempio}

	\begin{figure}
		\centering
		\includegraphics[width=\columnwidth]{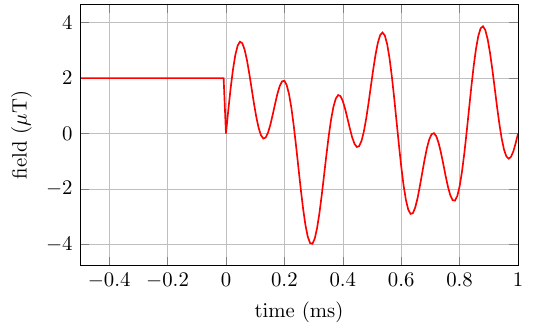}
		\caption{The test field-waveform consists of a constant value held for \SI{20}{ms} followed by a dual frequency signal, as described in Tab.~\ref{tab:testwf}. Here, only a short portion is shown to facilitate the $t=0$ transient visualization. }
		\label{fig:testwf}
	\end{figure}
	
	As a case study, we consider the requirements of an optical-pumping and spin-manipulation experiment~\cite{martin_prx_17}. The setup must maintain a DC magnetic field $B_p$ along a chosen direction (the $x$-axis) during an optical-pumping interval (from $-t_p$ to $0$). Subsequently, during the measurement interval ($0$ to $t_m$), the field must switch to a triaxial, time-dependent configuration. This consists of a static term $B_0$ along $z$ and two rotating terms in the $xy$ and $xz$ planes, each rotating at a distinct angular frequency. Thus, each field component is driven by a signal consisting of a static interval abruptly succeeded by either a single- or dual-frequency harmonic modulation.
	
	Accordingly, the magnetic field is expressed as
	\begin{equation}
		\vec B = \Bigl(B_{p,x}, 0 , 0\Bigr ) \qquad \text{for } -t_p < t < 0,
	\end{equation}
	and
	\begin{equation}
		\begin{split}
			\vec B = \Bigl ( & B_{1}\cos(\omega_1 t), \\
			& B_{1}\sin(\omega_1 t) + B_{2}\cos(\omega_2 t+\varphi), \\
			& B_{2}\sin(\omega_2 t+\varphi) + B_{0z} \Bigr ) \\
			& \text{for } 0 < t < t_m.
		\end{split}
	\end{equation}
	
	In our setup, the static field $\vec B_0$ and the dynamic ones $\vec B_p$, $\vec B_1$, and $\vec B_2$ are generated by distinct coil sets, each driven by an independent current source.  
	Since the static field $\vec B_0 = (0, 0, B_{0z})$ is provided by dedicated DC drivers configured to partially compensate the ambient magnetic field, the dynamic field that must be generated by the described system is
	
	\begin{equation}
		\vec B_\mathrm{dyn-p} = \Bigl(B_{p,x}, 0 , -B_{0,z}\Bigr) \qquad \text{for } -t_p < t < 0,
		\label{eq:Bdynp}
	\end{equation}
	and
	\begin{equation}
		\begin{split}
			\vec B_{\mathrm{dyn}\text{-}\mathrm{m}} = \Bigl( & B_{1}\cos(\omega_1 t), \\
			& B_{1}\sin(\omega_1 t) + B_{2}\cos(\omega_2 t+\varphi), \\
			& B_{2}\sin(\omega_2 t+\varphi) \Bigr) \\
			& \text{for } 0 < t < t_m.
		\end{split}
		\label{eq:Bdynm}
	\end{equation}
	
	The individual components of this time-dependent field must be controlled synchronously and with high precision, a demanding requirement because the desired waveforms exhibit discontinuities in amplitude or time derivatives~{\new\cite{haykin_book_96,oppenheim_book_99, rol_apl_20, duan_prl_26}}.
	
	From a physics standpoint, the fidelity requirement for the considered application is most stringent around $t=0$. Any glitch or overshoot at the switching instant perturbs the atomic ensemble, shifting the spin state away from the designed initial condition before the measurement begins. The limited bandwidth and compliance voltage of the amplifier stage make such artefacts unavoidable without pre-compensation, and suppressing them is the central objective of the method.

	\subsection{Results}

	\begin{table}[H]
	  \centering
		\begin{tabular}{llll}
			\hline \hline 
			\textbf{ } & \textbf{Description} & \textbf{Symbol} & \textbf{Value} \\
			\hline
			\multirow{3}{*}{Coil} & Inductance & $L_\mathrm{coil}$ & 8.5\,mH \\
			& Resistance & $R_\mathrm{coil}$ & $53\,\Omega$ \\
			& Calibration factor & $k_\mathrm{cal}$ & 200\,nT/mA \\
			\hline
			\multirow{3}{*}{RC Comp.} & Monitor resistance & $R_\mathrm{mon}$ & $100\,\Omega$ \\
			& Compens. resistance & $R_\mathrm{comp}$ & $100\,\Omega$ \\
			& Compens. capacitance & $C_\mathrm{comp}$ & $2.2\,\mu\text{F}$ \\
			\hline
			\multirow{2}{*}{Amplifier} & Gain & $G$ & 20 \\
			& Bandwidth & $BW$ & 0--20\,kHz \\
			\hline \hline
		\end{tabular}
		
		\caption{Coil characteristics and test circuit setup}
		\label{tab:setup}
	\end{table}

	We report representative results obtained for a single axis, with a setup characterized by the features summarized in Tab.~\ref{tab:setup}. To this end, we used the test waveform defined in Tab.~\ref{tab:testwf} and represented in Fig.~\ref{fig:testwf}, which shares features with field components described in eqs.~\ref{eq:Bdynp},~\ref{eq:Bdynm}. When generating triaxial fields, the precise time alignment between the designed and generated waveforms ensures the synchronicity of the triaxial Cartesian components, with the only requirement of time-matching among the DAC outputs.
	
	\begin{table}[H]
		\centering
		\begin{tabular}{llll}
			\hline \hline
			\textbf{} & Description & Symbol & Value \\
			\hline
			Sampling rate & Sa/s & $f_{DAQ}$ & $125\,$kHz \\
			\hline
			Static field & Duration & $\Delta t_\mathrm{stat}$ & $20\,$ms \\
			($t<0$)      & Level    & $B_\mathrm{static}$    & $2\,\mu$T \\
			\hline
			Dynamic field & Duration & $\Delta t_\mathrm{dyn}$ & $20\,$ms \\
			($t>0$)       & Amplitudes& $(A_1, A_2)$ & $(2,\,2)\,\mu$T \\
			& Frequencies& $(f_1, f_2)$ & $(2.5,\,6.0)\,$kHz  \\
			& Phases& $(\varphi_1, \varphi_2)$ & (0, 0) \\                          
			& Expression & \multicolumn{2}{l}{$\sum A_i \sin(2\pi f_i t+\varphi_i)$} \\
			\hline \hline
	\end{tabular}
		\caption{Test waveform design}
		\label{tab:testwf}
	\end{table}

	\begin{figure}
		\centering
		\includegraphics[width=\columnwidth]{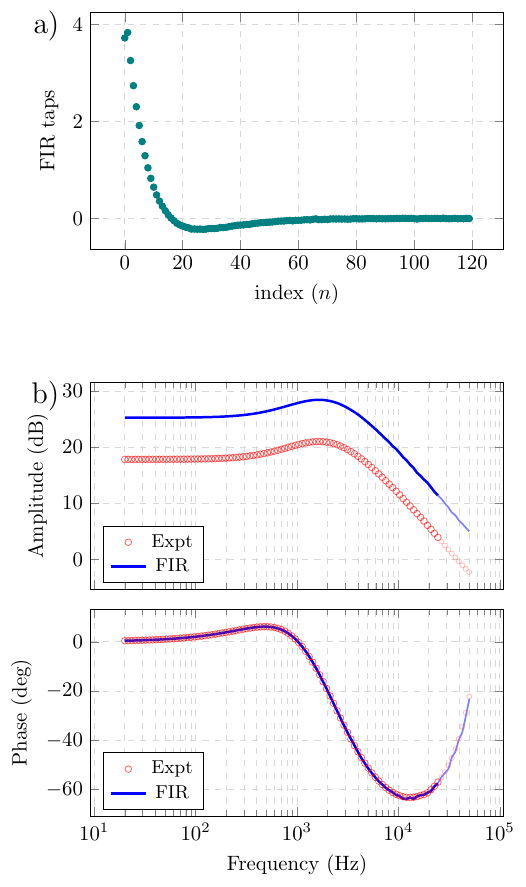}
		\caption{FIR taps (a) and corresponding bode plot (b). In b), the blue curves are evaluated with eq.\ref{eq:taps2bode}, while the red points are obtained experimentally, testing the system with harmonic signals whose frequency is scanned. The  8 dB displacement in the $|H|$ plots reflects the conversion factor from $V_\mathrm{mon}$ to {$B_m$}. {\new The lighter lines and marks correspond to frequencies above 25 kHz, outside the useful operating range}}
		\label{fig:fircoefficients}
	\end{figure}
	
	Fig.~\ref{fig:fircoefficients} shows the FIR tap set and the corresponding Bode-plot derived as
	
	\begin{equation}
		\begin{split}
			H(f) &= \sum_{n=0}^{N-1} h_n e^{-j 2\pi n \frac{f}{f_s}},  \\
			|H(f)|_{\text{dB}} &= 20 \log_{10} \left| H(f) \right|, \\
			\angle H(f) &=  \arg\{ H(f) \},
		\end{split}
		\label{eq:taps2bode}
	\end{equation}
	where $h_n$ represents the $n$-th FIR coefficient, $N$ is the total number of taps, and $f_s$ is the DAC sampling frequency.
	
\begin{figure}
		\centering
		\includegraphics[width=\columnwidth]{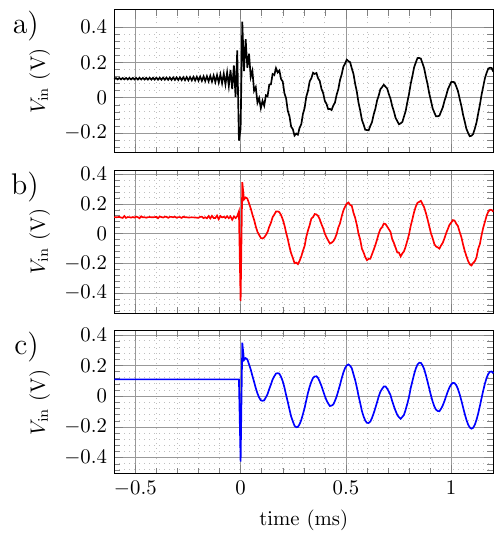}
		\caption{Input voltage required to produce the designed field waveform of Fig.~\ref{fig:testwf}. Only a short interval around the transition time is displayed. The black curve (a) represents $V_\mathrm{in}$ as estimated from the nominal circuit specifications (Tab.~\ref{tab:setup}); the red curve (b) is derived via FFT analysis of the measured system behavior; the blue curve (c) is obtained through the FIR identification method. }
	\label{fig:driving}
\end{figure}

	The input signal designed based on the FIR inversion is shown in Fig.~\ref{fig:driving}c for a limited interval around the transition time $t=0$, together with the corresponding traces evaluated based on 
	\begin{itemize}[noitemsep]		
		\item the nominal circuit specifications (Fig.~\ref{fig:driving}a)
		\item  a direct frequency domain characterization of the transfer function (Fig.~\ref{fig:driving}b).
	\end{itemize}
	{\new Taking case (c) as a reference, the RMS deviation of the input voltage over the plotted interval is about 38\,mV (5\,\% of the peak-to-peak excursion) for case (a) and 5\,mV (0.7\,\%) for case (b). Around the transient, signals (a) and (b) exhibit typical Gibbs oscillations, with frequencies of the order of 50\,kHz, which are absent (below the noise level) in case (c): in (a) they reach about 70\,mV in amplitude, last about 0.2\,ms before the transient and persist afterwards, whereas in (b) they are limited to about 17\,mV and 0.1\,ms and appear mostly before the transient.}

\begin{figure}
	\centering
	\includegraphics[width=\columnwidth]{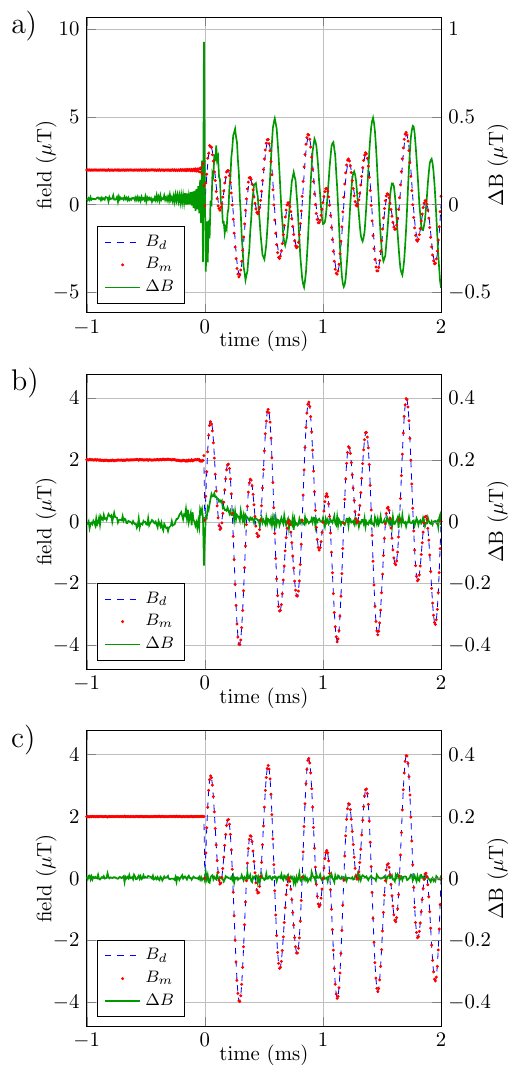}
	\caption{Designed and measured field around the $t_0$ transient. The plot a) is obtained precompensating the driving signal $V_\mathrm{in}$ based on the nominal values of the circuit. The plot b) uses a $V_\mathrm{in}$ calculated according to a FFT-based characterization of the circuit. The plot c) is obtained with the FIR identification and IFFT inversion. The blue line is the designed waveform, the red dots are the measured values, {\new and the green line is the error $\Delta B$, read on the right-hand axis (on the scale of the left-hand axis it appears magnified by a factor of ten).}}
	\label{fig:results}
\end{figure}

Finally, Fig.~\ref{fig:results}  shows the field waveforms generated by the system when driven by the input waveforms reported in Fig.~\ref{fig:driving}. 

It is worth noting that the relevance of waveform imperfections depends on the specific experiment and it must be evaluated in light of the specific requirements of the target application. 

The sensitivity of a quantum sensor --or, more generally, a quantum system-- to different forms of non-ideality (e.g. broadband glitches, narrow-band phase shifts,  long-term amplitude drifts, \dots) is strictly dependent on the specific Hamiltonian being engineered and the characteristic timescales of the coherent evolution. A single aggregate metric, such as an RMS error over the full waveform, could mask artifacts in critical intervals or other imperfections that can produce severe effects despite appearing with small amplitude. As a detailed analysis of diverse, specific experimental cases is beyond the scope of this work, we present time-domain error plots showing the residual field error in physical units, allowing the reader to judge its significance for a given application.
{\new Should aggregate metrics be required, they should be applied to specific data segments. For instance, Table~\ref{tab:errors}, based on the green traces in Fig.~\ref{fig:results}, reports RMS and maximum absolute deviation separately for the transient and the steady state.}

\begin{table}[htbp]
	\new
	\centering
	\begin{tabular}{l!{\hspace{1em}} S[table-format=1.3] S[table-format=1.3] !{\hspace{1em}} S[table-format=1.3] S[table-format=1.3]}
		\hline \hline
		
		& \multicolumn{2}{c}{Trans. $[-0.5,\,0.5]$\,ms\, \,} & \multicolumn{2}{c}{St. state $[2,\,3]$\,ms} \\
		\cmidrule(lr){2-3} \cmidrule(lr){4-5}
		Case & RMS & Max & RMS & Max \\
		\hline
		a) & 1.906 & 9.279 & 2.578 & 4.903 \\
		b) & 0.347 & 1.416 & 0.084 & 0.270 \\
		c) & 0.062 & 0.201 & 0.057 & 0.190 \\
		\hline \hline
	\end{tabular}
	\caption{RMS and maximum absolute deviation in $\mu$T of the error signals (green traces) shown in Fig.~\ref{fig:results}(a), (b) and (c), respectively, evaluated near the transient ($t\in[-0.5,\,0.5]$\,ms) and in the steady state ($t\in[2,\,3]$\,ms). In terms of RMS, the error in case (a) is about $31\times$ (transient) and $45\times$ (steady state) larger than in case (c), while in case (b) it is about $5.6\times$ (transient) and $1.5\times$ (steady state) larger than in case (c).}
	\label{tab:errors}	
\end{table}

\section{Discussion}
\label{sec:discussion}

The primary objective of this work was to develop and validate a versatile system capable of generating arbitrary triaxial magnetic waveforms with high fidelity and rapid reconfigurability. To overcome the non-ideal dynamics of the amplifier-coil network and meet diverse load specifications, we implemented a data-driven, two-step, feed-forward control strategy: a time-domain identification based on FIR filters optimized via WLS, followed by a frequency-domain inversion. This methodology ensures precise control and maximum versatility across a frequency range from DC to tens of kHz, enabling the hardware to adapt seamlessly to different experimental requirements.

The results of Sec.~\ref{sec:results} demonstrate that such FIR/WLS
identification strategy, combined with frequency-domain inversion,
effectively compensates for the non-ideal dynamics of the amplifier--coil
system. However, the inherent linearity of the compensation implies that
any unmodeled nonlinearity persists. In particular, nonlinearities can be
expected when the operational amplifier's output approaches its compliance
voltage limits. To this end, the peak values of the evaluated $V_\mathrm{in}$
are carefully monitored, and alerts are generated whenever such issues may
occur. {\new Likewise, the finite slew rate of the amplifier lies outside the
LTI model: the inverse filter demands steeper driving edges, and the Wiener
regularization, by bounding its high-frequency gain, helps keep the drive
within the linear regime, but cannot correct the distortion once slew
limiting occurs. Such a condition would manifest itself as a structured
residual in the error signal around the fastest transitions. In the waveform
considered here, the small residual error obtained around the transition
indicates that slew-rate limitation was not a significant issue.}

This hybrid workflow offers several practical advantages over model-based approaches relying on nominal circuit responses and over pure frequency-domain solutions:
\begin{itemize}[noitemsep]
	\item \textit{Accuracy in critical intervals:} The WLS weighting directs the model fit toward experimentally critical segments, suppressing transient artifacts at the field transitions. This flexibility is also exploited to assign zero weight to the initial $N$ samples so to exclude tap-loading transients.
	\item \textit{Elimination of precise model requirements:} The method removes the need for highly accurate knowledge of circuit element values~{\new\cite{merkel_rsi_19}}. The calibration signal $V_{\text{in},0}$ requires only a rough estimate of the system behavior; indeed, even neglecting the load reactance entirely in the initial guess yields an adequate starting point for the FIR identification.
	\item \textit{Physically relevant spectral coverage:} By deriving the calibration signal from the target waveform, the system inherently excites all frequency components present in the desired field. This ensures that the identified FIR filter is well-conditioned over the entire relevant bandwidth.
	\item \textit{Automatic compensation of unmodeled dynamics:} The data-driven nature of the filter automatically captures and compensates for (linear) non-idealities that are difficult to model analytically, such as the limited bandwidth of the power amplifier or other parasitic effects.
	\item \textit{Rapid adaptability and maintenance:} The method allows for on-the-fly recalibration to compensate for parameter drifts or hardware reconfigurations (e.g., dictated by varying waveform requirements) in only a few tens of seconds.
\end{itemize}

The core achievement of this procedure is the waveform fidelity obtained during demanding dynamic regimes, such as the abrupt transition from static to multi-frequency oscillating fields at $t=0$ considered in the reported example. As evidenced in Fig.~\ref{fig:results}, deriving the pre-compensated input voltage solely from nominal circuit parameters (a) or basic FFT-based characterizations (b) results in substantial transient artifacts and ringing. In contrast, evaluating the input waveform from the inverted FIR model almost entirely suppresses these distortions, minimizing the residual error. The WLS temporal weighting helps achieve this result, allowing the numerical identification to prioritize accuracy in specifically targeted critical intervals while disregarding initial tap-loading transients.

The two-layer design --comprising coarse hardware adjustment followed by software identification and pre-compensation-- renders the system both flexible and easy to maintain. Changing the coil, adapting to a different frequency range, or compensating for component drift requires only a new calibration run, which takes only a few tens of seconds on an ordinary computer. This is particularly valuable for experimental activities focused on the fine design and testing of dynamic fields for quantum state control with complex protocols, as demonstrated in the example considered in Sec.~\ref{subsec:esempio}.

That example motivates this work, highlighting the requirement for high-fidelity: artifacts at the $t=0$ switching instant perturb the atomic state and corrupt the initial conditions of the coherent evolution. The precise temporal control and inter-axis phase coherence provided by the proposed method are therefore directly relevant to experiments on effective Hamiltonian engineering and time-dependent Floquet couplings~\cite{bevilacqua_pra_22,fregosi_sr_23,fregosi_sr_26}.

The high fidelity achieved in current synthesis translates into magnetic field accuracy, provided that parasitic effects remain negligible, making the field proportional to the measured coil current. However, particularly at high frequencies, spurious effects due to conductive shields or parasitic capacitances may induce discrepancies between current and field dynamics. In such scenarios, the integration of in-situ calibration refinements, such as those described in \cite{bevilacqua_apdap_25}, would represent a natural extension to maintain precision and further broaden the method's applicability to non-ideal experimental environments

\section{Conclusions}
\label{sec:conclusions}

We have presented an integrated hardware and software system for the high-fidelity generation of arbitrary, triaxial magnetic waveforms from DC to tens of kilohertz. The method combines a reconfigurable amplifier stage with a data-driven identification and pre-compensation strategy based on FIR filtering and frequency-domain inversion. The key result is that waveform artefacts at demanding transitions --- which are unavoidable with nominal or simple FFT-based approaches --- are suppressed to levels comparable with the measurement noise floor.


The method requires no accurate knowledge of circuit parameters. Fast system recalibration can be performed after any hardware or waveform change, and concentrates its accuracy on the time intervals that are physically relevant for the experiment. These properties make it a practical and cost-effective tool for atomic-physics laboratories that require precise, rapidly reconfigurable magnetic field control.


\section*{Declarations}

\subsection*{Funding}
This research received no specific grant from any funding agency in the public, commercial, or not-for-profit sectors. The study was carried out as part of the ordinary institutional activities of the authors, supported by the Italian Ministry of University and Research (MUR).

\subsection*{Availability of Data and Material}
All data that support the findings of this study are included within the article.

\subsection*{Competing Interests}
The authors declare that they have no competing interests, whether financial, professional, or personal, that could influence the work reported in this paper.

\subsection*{Use of Artificial Intelligence}
During the preparation of this work, the author(s) used generative AI and AI-assisted technologies for language editing, and improving text readability. After using these tools, the authors reviewed and edited the whole text, and they take full responsibility for the content of the published article.

\subsection*{Authors' Contributions}
\begin{itemize}[leftmargin=*]
	\item \textbf{Conceptualization:}G.B., V.B., R.C.
	\item \textbf{Data curation:}V.B.
	\item \textbf{Formal analysis:}G.B., V.B.
	\item \textbf{Funding acquisition:} Not applicable.
	\item \textbf{Investigation:}G.B., V.B., R.C.
	\item \textbf{Methodology:}G.B., V.B., R.C.
	\item \textbf{Project administration:} Not applicable.
	\item \textbf{Resources:}G.B., V.B., R.C.
	\item \textbf{Software:}V.B.
	\item \textbf{Supervision:}V.B.
	\item \textbf{Validation:}G.B., V.B.
	\item \textbf{Visualization:}V.B.
	\item \textbf{Writing – original draft:} V.B.
	\item \textbf{Writing – review \& editing:}G.B., V.B., R.C.
\end{itemize}
All authors have read and agreed to the published version of the manuscript.

\bibliographystyle{ieeetr}
\bibliography{firbib.bib}
\end{document}